# Towards achieving strong coupling in 3D-cavity with solid state spin resonance

J-M. Le Floch,[1, 2, 3, a)] N. Delhote,[4] M. Aubourg,[4] V. Madrangeas,[4] D. Cros,[4] S. Castelletto,[5] and M.E. Tobar[2, 3]

[1)]*MOE Key Laboratory of Fundamental Physical Quantities Measurement, School of Physics, Huazhong University of Science and Technology, Wuhan 430074, Hubei, China*

[2)]*School of Physics, The University of Western Australia, Crawley, Western Australia 6009, Australia*

[3)]*ARC Centre of Excellence for Engineered Quantum Systems, Crawley, Western Australia 6009, Australia*

[4)]*XLIM, UMR CNRS 7252, Université de Limoges, 123 av. A. Thomas, 87060 Limoges Cedex, France*

[5)]*School of Aerospace, Mechanical and Manufacturing Engineering, RMIT University, Melbourne, Australia*

We investigate the microwave magnetic field confinement in several microwave 3D-cavities, using 3D finite-element analysis to determine the best design and achieve strong coupling between microwave resonant cavity photons and solid state spins. Specifically, we design cavities for achieving strong coupling of electromagnetic modes with an ensemble of nitrogen vacancy (NV) defects in diamond. We report here a novel and practical cavity design with a magnetic filling factor of up to 4 times (2 times higher collective coupling) than previously achieved using 1D superconducting cavities with small mode volume. In addition, we show that by using a double-split resonator cavity, it is possible to achieve up to 200 times better cooperative factor than the currently demonstrated with NV in diamond. These designs open up further opportunities for studying strong and ultra- strong coupling effects on spins in solids using alternative systems with a wider range of design parameters.

---

[a)] Electronic mail: jm_lefloch@hust.edu.cn, jeanmichel.lefloch@uwa.edu.au

**Strong coupling of paramagnetic spin defects with a photonic cavity is used in quantum computer architecture, to interface electrons spins with photons, facilitating their read-out and processing of quantum information. To achieve this, the combination of collective coupling of spins and cavity mode is more feasible, and offers a promising method. This is a relevant milestone to develop advanced quantum technology and to test fundamental physics principles.**

## I. INTRODUCTION

In recent years, paramagnetic spin defects in semiconductors and their quantum control at room temperature make them among the most relevant candidates for future scalable quantum computing[1]. For instance, spin defects in solid state hold promising applications spanning from ideal qubits[2] to unique magnetic resonance imaging and temperature probes[3,4]. One way to increase the coupling between systems is to use a large number of spins at once (collective coupling) coupled with either photonic or microwave cavities. The collective coupling, $g_c$, with the resonator mode can be enhanced by $g_c = g_s \sqrt{N}$[5,6], where $N$ is the number of identical two-level systems available (polarized number of spins). In our case, the spin ensemble can be treated as a simple harmonic oscillator[7]. The unique rigorous condition to reach strong coupling regime[8–11] is given by $g_c \gg \gamma_s \gg \kappa_c$, where $\kappa_c$ and $\gamma_s$ are the resonator and emitter damping rates respectively. Another way to describe the strength of the coupling between the spins and the cavity mode is given by the cooperative factor, where the strong coupling is measured by $C = g_c^2/(2\kappa_c\gamma_s) \gg 1$. Frequency splitting or anti-crossing may be seen from the reflected or transmitted signal. It does not necessarily mean that $g_c > \gamma_s$, and $g_c < \kappa_c$ and thus that light-matter coherent information can be transferred[10]. Relative to quantum-metrology, achieving strong coupling regime with spin ensemble, would give a higher sensitivity for detecting the signal absorbed or emitted by the spins[12]. Recently, high-Q superconducting coplanar waveguide (CPW) resonators were used as read-out of flux qubit[13], transmons[14,15], or spin defects[16–19]. The aim is to establish coherent transfer of information, by coupling two different quantum systems with a longer quantum coherence. The use of 3D superconducting cavities helps reaching longer coherence time but imposes many challenges in their realization as well as their future implementation. It has also been demonstrated that strong coupling to an ensemble of diamond nitrogen vacancy (NV⁻) spins is experimentally feasible [12,14,17,20,21].

In this paper we carefully determine the interaction between the electromagnetic field with an enhanced number of polarized spins[7] and the resonator topology. We give a detailed numerical study of 3D microwave cavity designs for optimizing the collective coupling between a spin ensemble and microwave photon. To validate our design, we use NV spin ensemble and compare against previous 1D superconducting planar cavities[17,21].

## II. PROOF OF PRINCIPLE ANALYSIS

### A. Spin ensemble description

Among all atomic-like solid state systems, we here focus on a well-known spin defect in diamond, the negatively charged state of the nitrogen vacancy center (NV⁻). NV has recently attracted a great interest, due to the unique properties of its ground state spin. The latter is an individually addressable solid state quantum bit at room temperature[13,19], with a very long coherence time[15]. Also, quantum control using an optical read-out[13,22] can be implemented. NV center in diamond is constituted by a substitutional nitrogen atom close to a carbon vacancy having trapped an additional electron (Fig.1); its electronic ground state spin S =1, with the state $m_S=0$ and $m_S=\pm 1$, is separated by 2.87GHz at zero magnetic field[23]. In the case of NV⁻, $m_0 = g_{NV} \mu_B$ is the electron magnetic moment, where $\mu_B$ is the electron Bohr magneton, $g_{NV}$ = 2.0028 (S=1) is the NV Landé factor, owing to a ground state electron spin Hamiltonian given by $H_{NV} = m_0 \mathbf{B}.\mathbf{S} + \mathbf{S}.\mathbf{D}.\mathbf{S}$. **B** is an external DC magnetic field, **D** the zero field splitting for the axial component along NV axis ( $\frac{D}{h}$ = 2.877GHz ). The hyperfine coupling to the $^{14}$N or $^{13}$C nuclear spin is neglected for simplicity.

### B. Electromagnetic wave distribution in cavities

#### 1. Cavity to spin interaction

A microwave cavity consists of a metallic enclosure that confines electromagnetic fields in the microwave region of the spectrum. The structure is either unloaded or loaded with one or more dielectric materials. For single spin coupling, the coupling strength depends on the cavity mode volume, whereas, for ensemble of spins, the collective coupling depends only on the filling factor and the number of polarized spins[14]. The coupling interaction between spins and the microwave.

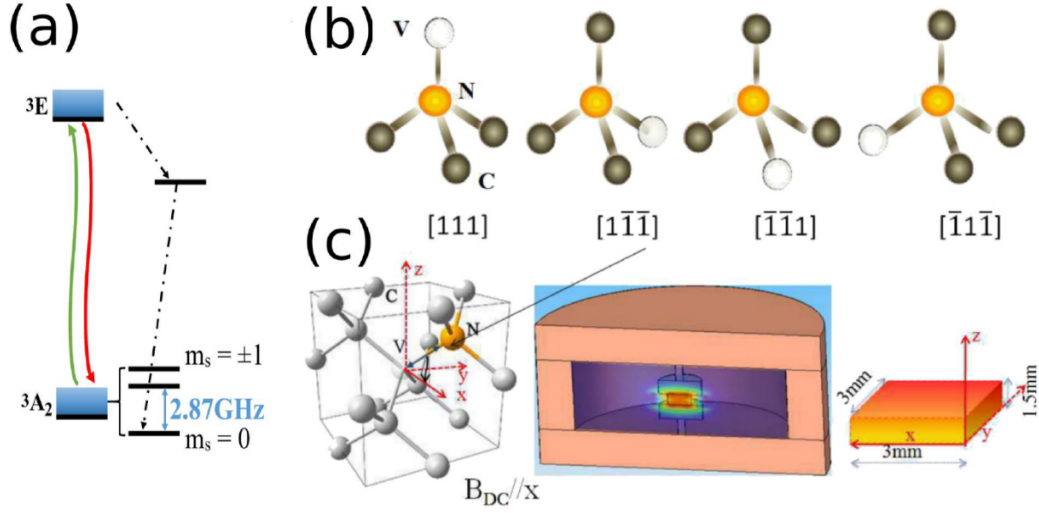

FIG. 1: (a)Simplified energy diagram of NV center in diamond, showing the ground state spin level splitting and the excited state radiative and non-radiative transitions. (b)NV axis, four possible orientations, referred to the diamond crystallographic axis (c) NV shown with its crystallographic location in a diamond cell (left) along one of the crystallographic direction <111>, while z indicated the main c-axis [001]. We consider that the bulk diamond (3x3x1.5 mm) is placed in the cavity (example shown of a double-split resonator) with the correct orientation to achieve only two sub-ensemble of NV defects aligned at 45° with the DC applied magnetic $B_{DC}$ field.

field of the cavity can be described as a basic spin harmonic oscillator. The collective coupling strength can be expressed from the basic spin harmonic oscillator approximation, as given by[14],

$$g_c = \frac{m_0}{2}\sqrt{\frac{\rho\mu_0\omega_c p_m}{\hbar}}, \qquad (1)$$

where $\rho$ is the number of spins per unit volume, $\omega_c$ the cavity resonance frequency, and $\mu_0$ the vacuum permeability.

The cavity magnetic filling factor, $p_m$, describes the AC-magnetic field confinement in a particular volume of spins.

$$p_m = \frac{W_{m_{material}}}{W_{m_{Total}}} = \frac{\oiiint_{V_{material}} \mu_r H.H^* dv}{\oiiint_{V_{Total}} \mu_r(v) H.H^* dv}, \qquad (2)$$

with $\mu_r$ the relative permeability, which is in our case equals to 1. $p_m$ enters in increasing the collective spin coupling $g_c$, see Eq.1, as the cavity electromagnetic mode must be highly confined within the spins volume density. We also assume that the spins volume density is uniform within the volume, thus it is not entering in the magnetic filling factor calculations.

*2. Cavity parameters determination*

We present here topologies that allow specifically the insertion of a bulk diamond sample rich in NV-spins, where the magnetic field confinement ($p_m$) is the highest. The dielectrics loss (diamond and if applicable, the dielectric host) contribute to the resonator Q-factor. This dielectric loss is characterized by the microwave electric field confinement ($p_e$) and its intrinsic loss mechanism ($tan\,\delta$).

$$p_e = \frac{W_{e_{material}}}{W_{e_{Total}}} = \frac{\oiiint_{V_{material}} \epsilon_{material} E.E^* dv}{\oiiint_{V_{Total}} \epsilon(v) E.E^* dv}, \qquad (3)$$

where $\varepsilon$ is the relative dielectric permittivity.

The dielectric loss dominates the cavity photons damping rate. Eventually, it drives the information encoded in the spin. The resonator damping rate $\kappa_c$ can be measured from the cavity resonance bandwidth:

$$Q_0 = \frac{2\pi\nu_c}{\kappa_c} = \frac{\omega_c}{\kappa_c}, \qquad (4)$$

where $\nu_c$ is the resonance frequency, $\omega_c$ is the pulse cycle of the resonant cavity, and $Q_0$ is the unloaded Q-factor of the cavity, given by:

$$Q_0^{-1} = Q_{met}^{-1} + Q_{diel_{Diamond}}^{-1} + Q_{diel_{host}}^{-1}, \qquad (5)$$

where $Q_{met}$ and $Q_{diel}$ correspond to the Q-factor of the metallic walls of the cavity and the Q-factor of the dielectrics (diamond and host) loaded into the cavity, respectively[24]. We also focus on reducing the limitations due to metallic losses estimated as:

$$Q_{met}^{-1} = \frac{R_{surface}}{GF}, \qquad (6)$$

$R_{surface}$ is related to the skin depth and metal conductivity of the cavity. And the geometric factor (GF) represents the microwave magnetic field distribution along the cavity surface.
Finally,

$$Q_{diel}^{-1} = p_e \tan \delta, \qquad (7)$$

where *tan δ* is the loss term of the dielectric material (diamond or host).

## 3. Simulation description

In our simulations, we assume that the contribution from the spin average orientation with respect to DC applied **B**-field is the same in all studied cavities design. In fact, in irradiation produced NV centers in diamond, care must be used to position the diamond to achieve two sub-ensemble of NV defects aligned at 45 degrees with the DC applied magnetic ($B_{DC}$) field (Fig.1). This condition is design independent as it can always be achieved by being just a matter of mechanical fitting of the cavity-diamond inside the cryo-refrigerator and within the B-field homogeneity. Actually, the **B**-field homogeneity, using conventional magnet, is at least 1cm DSV (Diameter Sphere Volume), corresponding to a volume big enough to fully cover the diamond sample. The rotation of the cavity, to align a particular NV-center orientation with the DC-applied magnetic field, doesn't modify its electromagnetic properties. Therefore, we assume that the same cavity- diamond positioning is used for all the designed cavities to optimize the number of the coupled spin subset and all the studied cavities rely on the same probed spin densities. In the case of bulk diamond, where preferentially aligned NV spins are grown, all the spins in the material could be effectively coupled to the cavity modes[25].
We only consider cavity modes with an AC-magnetic field propagating through a diamond sample. In this case, this is a fair assumption as the field occupies the volume of the whole sample. All simulations of the different structures presented here are conducted using the materials properties at 4K.

The interest to be as low temperature as possible, in the mK regime, is to maximize the chance to detect strong coupling by lowering down the microwave noise and optimizing the cavity Q-factor. The aim, in this investigation, is to find the resonator with the smallest bandwidth and the highest AC-magnetic field confinement for enhancing the magnetic-dipole interaction of the defects. Thus, in the microwave regime, to match both conditions, it is necessary to use single crystal low-loss dielectrics at low temperature. In addition, the sub-Kelvin temperature experiment ensures that the spins are not thermally polarized. We used a simulation software based on finite element analysis, developed for the past 20 years at XLIM institute and specifically optimized for microwave resonators.

The simulated resonance is 2.87GHz±20MHz, matching the microwave driving frequency of the NV-center from $m_s$=0 to $m_s$=±1. The magnetic $p_m$ (Eq.2) and electric $p_e$ (Eq.3) field confinements, relative to the cavity mode resonances into the NV-center defect in diamond, are calculated to determine the proportion of fields being confined in the diamond sample.

## 4. *Cavity based on waveguide design*

Here, we first focused on waveguide-based technologies, to confine the AC-magnetic field. We compare cavity topologies by the extend of the magnetic field confinement in the diamond sample. Different topologies of waveguides were used in various experiments and fields of research, such as gyrotron[26], wedge, ridge, and L-shape, including the use of septas[27]. These were adapted into cavities by closing both extrema of the guide with a metallic plane.

All these are interesting for this experiment as magnetic field constraint, frequency and shape are important parameters. They highly confine the electromagnetic field into their centers. Quarter-wavelength ridge long and half-wavelength L-shape waveguides use a high capacitance effect to reduce the resonance frequency. This capacitance is generated by a high-electric field between the large inserted metallic pieces (bar or L-shape). The capacitance enables separating significantly areas where the electric and magnetic fields are confined.

Wedges and septas introduce perturbations of the electromagnetic field pattern propagating in the cavity. These perturbations allow a better field confinement into a particular area. This technique suffers from lowering the Q-factor due to large metallic losses. The aforementioned cavities are illustrated in Fig.2. Such cavities are big at this resonance frequency, resulting in a very low AC-magnetic field confinement into the diamond, and thus, lowering down the sensitivity to detect strong coupling.

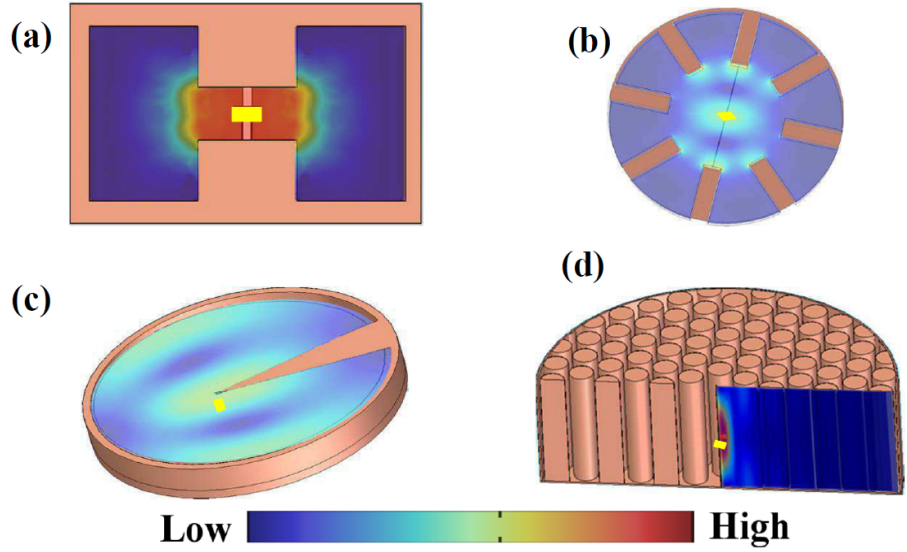

FIG. 2: Magnetic field density plots of the waveguide based technology, (a) ridge, (b) septa, (c) wedge, (d) gyrotron. Due to scale ratio between the cavity and the diamond sample, the diamond (yellow square) is enhanced to illustrate the place where it is meant to be located.

### C. High-confinement microwave cavities

Compared to 1D cavities, a wider range of possible design and realization parameters are pro- posed to reach strong coupling towards ultra-strong coupling regime[28]. These enable further fundamental studies on cavity quantum electrodynamics due to their design flexibility and permit to tailor different spins systems.

#### 1. High-Q dielectric resonators

We previously designed 3D cavities such as Transverse Electric (TE)[29,30], lumped reentrant[27,31–34], Fabry-Perot (FP)[35–38] and Whispering Gallery Mode (WGM)[39–41] resonators. High Q-factor cavities, with very narrow linewidth, are ideal for strong coupling detection. For this purpose, we also consider other cavities, such as high-Q dielectric Bragg, photonic bandgap resonators and whispering gallery cylindrical and spherical resonators (see Fig.3), which however provide very low AC-magnetic field confinement (less than $10^{-3}$).
At such a low frequency, WGM, Bragg, and band gap resonators are limited by their sizes compared to diamond dimensions. This prevents us from fulfilling the strong coupling conditions. In addition, to fully confine a high magnetic field into the diamond sample with WGM, it is necessary to excite a large azimuthal mode number, which then reduces the maximum possible confinement per field maxima (less than 0.03).

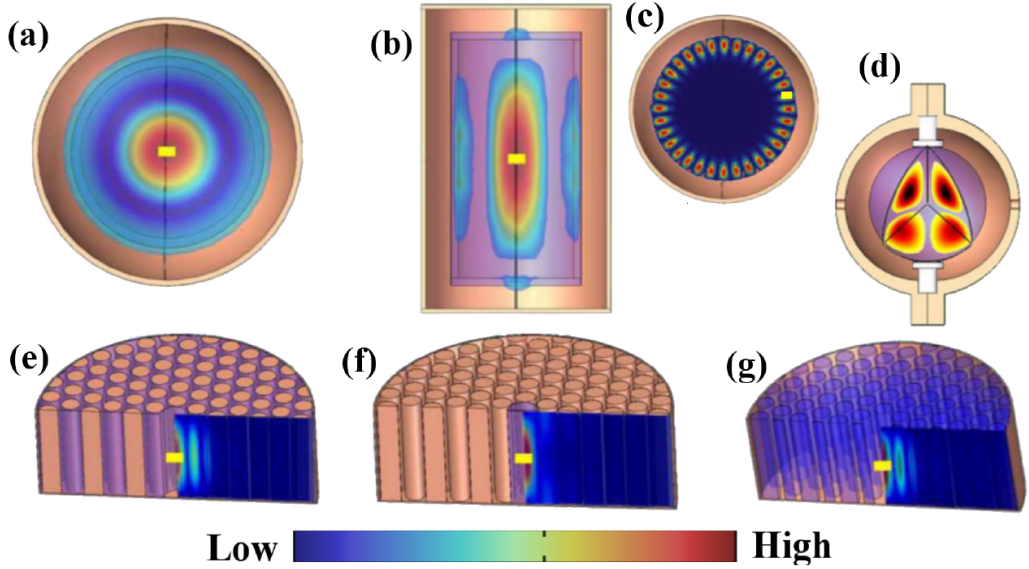

FIG. 3: Magnetic field density plots of (a), and (b) Bragg resonators, (c) and (d) whispering gallery modes both cylindrical and spherical symmetries, (e), (f) and (g) photonic band gap resonators[42–46]. For these different topologies, low loss dielectrics (purple color) have been used[47–49]. Due to scale ratio between the cavity and the diamond sample sizes, the diamond (yellow square) is enhanced to illustrate the place where it is meant to be located.

*2. Cylindrical symmetry reentrant cavities*

Cylindrical reentrant cavities are unique 3D-structures, where their electric and magnetic fields are located in separate parts of the cavity. Reentrant cavities, illustrated in Fig.4, offer the advantage of a small size with a high confinement of magnetic field around the central post[31–33] which imposes to drill a concentric hole into the sample to fit the cavity.
A double-post or periodic reentrant cavity can be built[27,34] to place the sample in a high confined region of the cavity. With these topologies, the machining of the sample is not necessary but they exhibit low-Qs. As a consequence, we have studied other high-Q dielectric loaded cavities based on transverse electric (TE) modes.

### 3. *Double-split mode cavities*

When low-loss dielectric materials are inserted in transverse electric (TE) mode cavities, the cavity size can then be reduced and the AC-magnetic field confinement increases[30,50–52]. One cavity design can respond to these criteria, the double-split cavity. It consists of a metallic enclosure loaded with two dielectric discs upside down, creating a gap in between for another dielectric material to be inserted (Fig.5 and 6). Our proposed double-split cavity is composed of a diamond sample sandwiched between two discs of $TiO_2$. The whole is inserted into a copper enclosure. The resonator mode is $TE_{0,1,\delta}$. The large dielectric permittivity of $TiO_2$ confines the field around the sample and allows a small size cavity for a 2.87GHz resonant frequency. $TiO_2$ low-microwave loss enables high-Qs, about 40 times higher than the reentrant cavity, with the same magnetic field

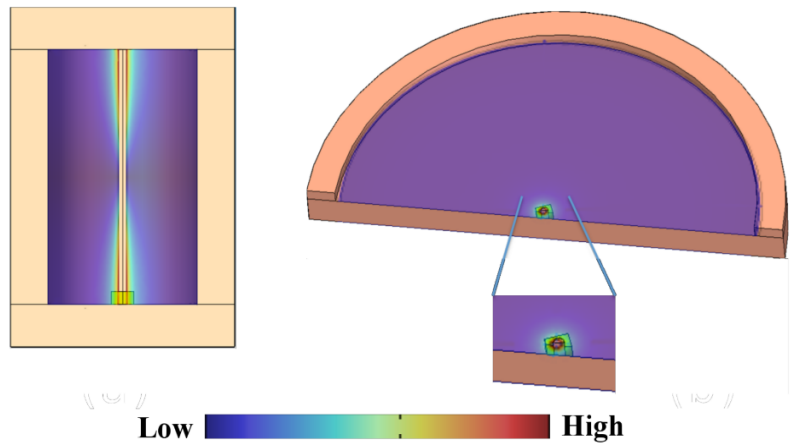

**Low**     **High**

FIG. 4: Magnetic field density plots of (a)coaxial reentrant, and (b) reentrant cavity. The diamond (yellow square) on figure (a) is enhanced to show the location. On figure (b) only the contour is shown to show the magnetic field.

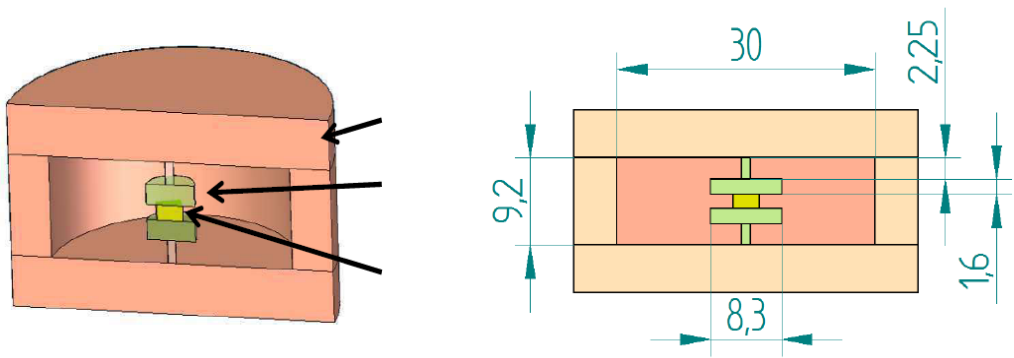

FIG. 5: (a) and (b) illustrate the double split-post resonator design, where 8mm-diameter c-axis parallel $TiO_2$ discs confine the field into the diamond sample and hold tight in position the diamond sample.

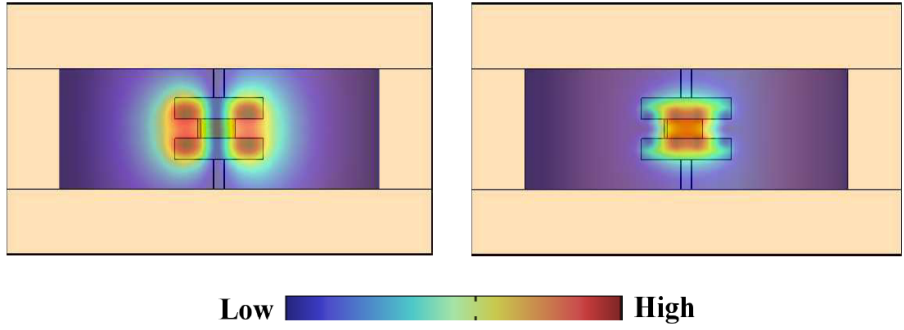

FIG. 6: (a) and (b) illustrate the density plots of the electric field and magnetic field respectively. The color coding from blue to red corresponds to low to high intensity field.

confinement inside the sample (0.2).

For achieving strong coupling, we need to design a small volume cavity with a relatively high-Q and a very high AC-magnetic field confinement, where a dielectric host is not necessary to avoid uncertain coupling conditions.

### 4. *Development of a hybrid cavity*

The hybrid cavity design is a combination of topologies and aims at reducing cavity size and preventing the use of an additional dielectric, while maintaining a high AC-magnetic field confinement. For that particular purpose, the waveguide-based technology brings techniques to manipulate the field pattern and direction. The reentrant cavity does not require dielectric and is of small size. It also enables a high-field confinement. Combining the reentrant cavity with the waveguide, wedge, and septa topologies[53–55], we finally design an unloaded hybrid cavity (Fig.7 and 8).

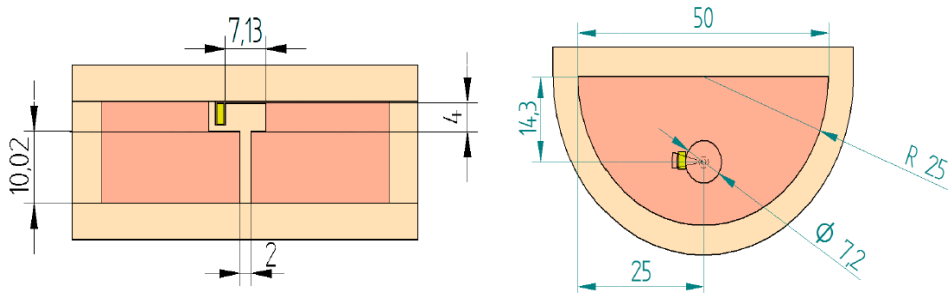

FIG. 7: (a) and (b) illustrates the hybrid cavity design with the inserted diamond sample.

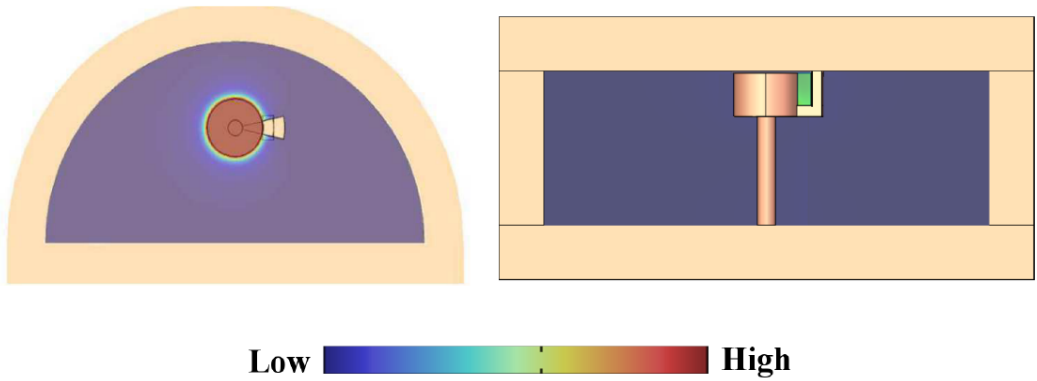

FIG. 8: (a) and (b) represent both the density plots of electric field (confined within the gap formed between the post and the top lid of the cavity) and the magnetic field (highly confined into the diamond sample). The color coding from blue to red corresponds to low to high intensity field.

To reduce the size of the cavity we use the electric symmetry properties of a transverse electric cavity mode through a pi-wedge design. Then the insertion of a pin (septa) through the height of the cavity modifies the wave propagation and creates a short-circuit. It automatically draws currents and intensifies the magnetic field. Finally, to reduce the frequency of the cavity down to 2.87GHz, a large capacitive effect has to be created such as a reentrant cavity. The mushroom shape of the pin has no other function than reducing the metallic losses of the cavity. Due to the constructive combination of different technologies, a maximum magnetic field confinement into the diamond (0.47) can be achieved.

III. VALIDATION OF MODEL AND ESTIMATION OF PERFORMANCE

In this section, we provide the results for the most promising cavities loaded with the NV diamond sample: TE double split resonator and the hybrid cavity, shown in Fig.5,6,7 and 8, respectively. 1D CPW cavities[21] are used as reference to validate the calculations. The results of the simulations are summarized in Table I, where only the best results for a same mode are reported with a comparison to the 1D CPW resonator[21]. Two structures have clear interesting features for achieving strong coupling. First, the double split-post resonator, shown in Fig.5 and 6, has a very high-Q factor with a minimum of magnetic field confinement of 0.2 into the diamond sample. This design requires to have a large number of NVs. Secondly, the unloaded hybrid cavity, presented in Fig.7 and 8, allows confining the magnetic field to 0.47 which is the highest field concentrated enabling structure. This means less NVs are necessary to achieve strong coupling. In such a structure, ultra-strong coupling could then be considered.

TABLE I: Summary of designed and computational parameters to determine the performance of each electromagnetic mode and cavities ($p_m$, $p_e$, $Q_0$) and also predict the coupling strength $g_c$ and the cooperative factor C. Mode denotes the electromagnetic mode structure the calculations are related to. We mostly used as a host for diamond[56], $Al_2O_3$[47,57] (sapphire), $TiO_2$[48] (rutile) and fused silica[58]. We use copper[59] for the cavity walls (Rs=5.77mΩ). All simulations are conducted at 4K and the resonance frequency is 2.87GHz±20MHz.

| Mode | $p_m$ ×10³ | $p_e$ ×10³ | $Q_0$ | $g_c$ (MHz) | V (cm³) | C |
|---|---|---|---|---|---|---|
| **Double-split resonator- Fig.5,6** | | | | | | |
| TE[48,50] | 84 | 0.63 | 127,000 | 43 | 14 | 348 |
| TE*[48,52] | 207 | 0.02 | 300,000 | 68 | 106 | 2,027 |
| **Reentrant cavities- Fig.4** | | | | | | |
| TM[33] | 100 | 0.35 | 1,000 | 47 | 36 | 3.3 |
| TM**[31,33] | 230 | 0.26 | 1,250 | 71.5 | 36 | 9.4 |
| TM[34,60] | 270 | 0.27 | 500 | 77.5 | 36 | 4.4 |
| **Unloaded hybrid cavities (new topology)- Fig.7,8** | | | | | | |
| copper | 471 | 49 | 500 | 102 | 14 | 7.7 |
| with Nb post | 469 | 49 | 1,000 | 102 | 14 | 15.3 |
| **Coplanar resonator-small volume** | | | | | | |
| [21] | 119 | 215 | 1,905 | 51 | 0.2 | 7.4 |

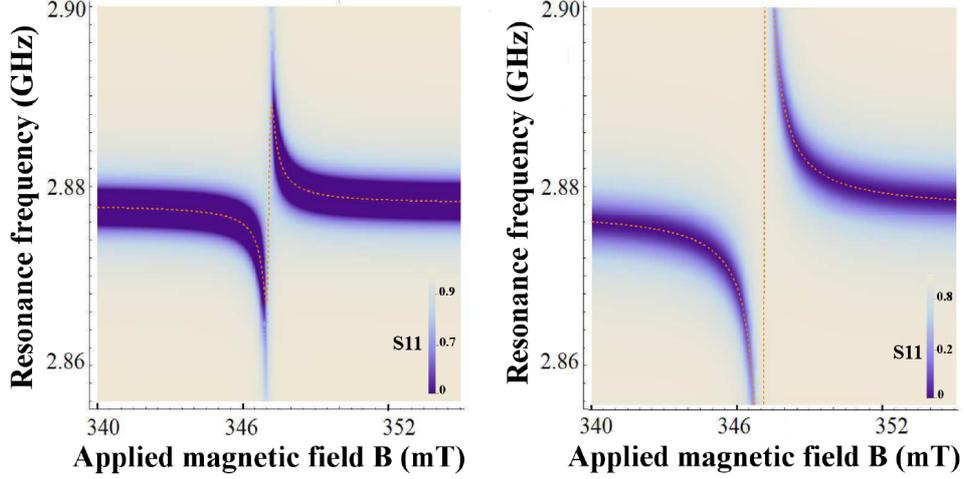

FIG. 9: $|S_{11}|^2$ spectrum for (a) CPW resonator, (b) hybrid cavity, our new topology with Niobium (Nb) post to reduce losses. The dashed lines correspond to Eq.9.

The linewidth of the spins compared with the cavity bandwidth has to be greater, as this cavity suffers from being low-Q. The cavity transmission mode showing frequency splitting once the cavity mode couples with the spin ensemble is given by

$$|S_{11}|^2 = \left|1 + \frac{\kappa_e}{i(\omega - \omega_c) - \kappa + g_c^2/(i\Delta - \gamma)}\right|^2, \quad (8)$$

where $\kappa_e = \alpha\,\kappa_c$ is the external loss.

In Fig.9, we illustrate the computed reflection spectra (from Eq.8 and Table I) of an ensemble of NV-spins for the case of a coplanar resonator with small volume, corresponding to the reported measurements[21] and for our hybrid cavity design with large volume. Even though the latter exhibits a lower Q-factor, the magnetic field confinement is higher, thus increasing the coupling strength significantly. We assume the density of spins $\rho = 1.2 \times 10^6 \mu m^{-3}$, which is typically achieved in HPHT diamond, after high energy electron/neutron irradiation or ion implantation[17,21]. This value is usually measured by using confocal microscopy, and comparing the photo-luminescence (PL) of ensemble to a single defect. $\gamma_S/2\pi$ is the FWHM linewidth of the ESR lines of NV spins ensemble, proportional to the inverse of the spin dephasing time or phase relaxation time ($T_2^*$). The ESR linewidth of the NV-center in diamond is sensitive to temperature and irradiation dose. For our simulations, we assume $\gamma_S/2\pi \approx 3$ MHz as we operate at low temperature. This is the expected broadening due to dipolar interactions with the neighboring $^{14}$N electronic spin (S=1/2). Typically, in 100 ppm HPHT diamond we may expect $\gamma_s \approx 18.84$ MHz.

A model with two-coupled oscillators describes the change in cavity frequency $\omega$ and cavity half-width once the cavity couples with spins in presence of a DC magnetic field $B$:

$$\omega = \omega_c + g_c^2 \Delta / (\Delta^2 + \gamma_s^2), \qquad (9)$$

where $\Delta = m_o \times (B - B_r)/\hbar$ is the field detuning from the resonant DC magnetic field $B_r$.

## IV. DISCUSSION

In summary, we have presented 3D-cavities for achieving strong coupling at the exact transition of the NV-diamond ground state spin resonance; we verified the modes through numerical finite-element modeling and validated our models comparing the obtained designed cavities parameters with some experimental realizations[17,30,34]. We also introduced a variety of topologies. We designed novel structures based on an unloaded hybrid cavity that confines four times the magnetic field into the diamond sample. It can provide twice the cooperative coupling, compared to small volume coplanar waveguides (CPW). This new type of cavity exhibits the highest magnetic field confinement in microwave cavities. In addition, this new cavity design can have potential applications in testing fundamental physics, such as paraphoton detection[61]. From our investigation, results show that strong coupling could also be achieved by a double-post cavity, where the collective coupling strength can be of the same order of the CPW but with 200 times better cooperative factor. NV centers also offer access to an additional optical transition, which can be of interest in 3D cavities with optical access to transfer quantum information from optical to microwave. The cavity designs allowing strong coupling and presented in this paper, cannot provide a direct optical access, further investigation would be required.

3D-microwave cavities presented in this paper have a wide range of design parameters, they can also be applied to emerging similar qubits in silicon carbide[62–64] or $Ce^{3+}$ in YAG[65]. The best cavities investigated here will certainly help by leading to even stronger couplings using other spins available in the solid state with much narrower linewidths than NV centers, enabling new physics tests. If we were to use this particular design principle at different temperatures, the magnetic field confinement would remain identical within 5%. Conversely, at different temperatures, the Q-factor of these cavities will strongly depend on the metallic enclosure conductivity or on its dielectric loss mechanism in the case of double-split resonators. Microwave temperature noise will increase with increasing temperatures, limiting the use of microwave cavities at room temperature.

**ACKNOWLEDGMENTS**

This research is jointly supported under the Australian Research Council funding scheme: Laureate Fellowship (project number FL0992016), and Centre of Excellence Engineered Quantum Systems (project number CE110001013), the French Research Agency (CNRS), Labex Sigma-Lim (No. ANR-10-LABX-0074-01). The authors also thank le Conseil Régional du Limousin, cluster de calcul CALI (Calcul en Limousin) and the UWA Research Collaboration Award scheme.

**REFERENCES**

[1] D. D. Awschalom, L. C. Bassett, A. S. Dzurak, E. L. Hu, and J. R. Petta, "Quantum spintronics: Engineering and manipulating atom-like spins in semiconductors," Science **339**, 1174–1179 (2013).

[2] W. Pfaff, B. J. Hensen, H. Bernien, S. B. van Dam, M. S. Blok, T. H. Taminiau, M. J. Tiggelman, R. N. Schouten, M. Markham, D. J. Twitchen, and R. Hanson, "Unconditional quantum teleportation between distant solid-state quantum bits," Science **345**, 532–535 (2014).

[3] J. Taylor, P. Cappellaro, L. Childress, L. Jiang, D. Budker, P. Hemmer, A. Yacoby, R. Walsworth, and M. Lukin, "High-sensitivity diamond magnetometer with nanoscale resolution," Nature Physics **4**, 810–816 (2008).

[4] G. Kucsko, P. Maurer, N. Yao, M. Kubo, H. Noh, P. Lo, H. Park, and M. Lukin, "Nanometre-scale thermometry in a living cell," Nature **500**, 54–58 (2013).

[5] F. Brennecke, T. Donner, S. Ritter, T. Bourdel, M. Köhl, and T. Esslinger, "Cavity qed with a bose–einstein condensate," Nature **450**, 268–271 (2007).

[6] Y. Colombe, T. Steinmetz, G. Dubois, F. Linke, D. Hunger, and J. Reichel, "Strong atom–field coupling for bose–einstein condensates in an optical cavity on a chip," Nature **450**, 272–276 (2007).

[7] A. Imamoğlu, "Cavity qed based on collective magnetic dipole coupling: spin ensembles as hybrid two-level systems," Physical review letters **102**, 083602 (2009).

[8] R. Schoelkopf and S. Girvin, "Wiring up quantum systems," Nature **451**, 664–669 (2008).

[9] J. M. Fink, *Quantum nonlinearities in strong coupling circuit QED*, Ph.D. thesis, University of Vienna (2010).

[10] E. Abe, H. Wu, A. Ardavan, and J. J. Morton, "Electron spin ensemble strongly coupled to three-dimensional microwave cavity," Applied Physics Letters **98**, 251108 (2011).

[11] G. Boero, G. Gualco, R. Lisowski, J. Anders, D. Suter, and J. Brugger, "Room temperature strong coupling between a microwave oscillator and an ensemble of electron spins," Journal of Magnetic Resonance **231**, 133–140 (2013).

[12] Y. Kubo, I. Diniz, C. Grezes, T. Umeda, J. Isoya, H. Sumiya, T. Yamamoto, H. Abe, S. Onoda, T. Ohshima, *et al.*, "Electron spin resonance detected by a superconducting qubit," Physical Review B **86**, 064514 (2012).

[13] P. Neumann, R. Kolesov, B. Naydenov, J. Beck, F. Rempp, M. Steiner, V. Jacques, G. Balasubramanian, M. Markham, D. Twitchen, *et al.*, "Quantum register based on coupled electron spins in a room-temperature solid," Nature Physics **6**, 249–253 (2010).

[14] D. Schuster, A. Sears, E. Ginossar, L. DiCarlo, L. Frunzio, J. Morton, H. Wu, G. Briggs, B. Buckley, D. Awschalom, *et al.*, "High-cooperativity coupling of electron-spin ensembles to supercon- ducting cavities," Physical review letters **105**, 140501 (2010).

[15] G. Balasubramanian, P. Neumann, D. Twitchen, M. Markham, R. Kolesov, N. Mizuochi, J. Isoya, J. Achard, J. Beck, J. Tissler, *et al.*, "Ultralong spin coherence time in isotopically engineered diamond," Nature materials **8**, 383–387 (2009).

[16] J. Healey, T. Lindström, M. Colclough, C. Muirhead, and A. Y. Tzalenchuk, "Magnetic field tuning of coplanar waveguide resonators," Applied Physics Letters **93**, 043513 (2008).

[17] X. Zhu, S. Saito, A. Kemp, K. Kakuyanagi, S.-i. Karimoto, H. Nakano, W. J. Munro, Y. Tokura, M. S. Everitt, K. Nemoto, *et al.*, "Coherent coupling of a superconducting flux qubit to an electron spin ensemble in diamond," Nature **478**, 221–224 (2011).

[18] H. Paik, D. I. Schuster, L. S. Bishop, G. Kirchmair, G. Catelani, A. P. Sears, B. R. Johnson, M. J. Reagor, L. Frunzio, L. I. Glazman, S. M. Girvin, M. H. Devoret, and R. J. Schoelkopf, "Observation of high coherence in josephson junction qubits measured in a three-dimensional circuit qed architecture," Phys. Rev. Lett. **107**, 240501 (2011).

[19] N. Bar-Gill, L. M. Pham, A. Jarmola, D. Budker, and R. L. Walsworth, "Solid-state electronic spin coherence time approaching one second," Nature communications **4**, 1743 (2013).

[20] R. Amsüss, C. Koller, T. Nöbauer, S. Putz, S. Rotter, K. Sandner, S. Schneider, M. Schramböck, G. Steinhauser, H. Ritsch, *et al.*, "Cavity qed with magnetically coupled collective spin states," Physical review letters **107**, 060502 (2011).

[21] Y. Kubo, F. Ong, P. Bertet, D. Vion, V. Jacques, D. Zheng, A. Dréau, J.-F. Roch, A. Auffèves, F. Jelezko, *et al.*, "Strong coupling of a spin ensemble to a superconducting resonator," Physical review letters **105**, 140502 (2010).

[22] L. Childress, M. G. Dutt, J. Taylor, A. Zibrov, F. Jelezko, J. Wrachtrup, P. Hemmer, and M. Lukin, "Coherent dynamics of coupled electron and nuclear spin qubits in diamond," Sci- ence **314**, 281–285 (2006).

[23] M. Doherty, F. Dolde, H. Fedder, F. Jelezko, J. Wrachtrup, N. Manson, and L. Hollenberg, "Theory of the ground-state spin of the nv- center in diamond," Physical Review B **85**, 205203 (2012).

[24] J.-M. Le Floch, Y. Fan, G. Humbert, Q. Shan, D. Férachou, R. Bara-Maillet, M. Aubourg, J. G. Hartnett, V. Madrangeas, D. Cros, *et al.*, "Invited article: Dielectric material characterization techniques and designs of high-q resonators for applications from micro to millimeter-waves frequencies applicable at room and cryogenic temperatures," Review of Scientific Instruments **85**, 031301 (2014).

[25] J. Michl, T. Teraji, S. Zaiser, I. Jakobi, G. Waldherr, F. Dolde, P. Neumann, M. W. Doherty, M. B. Manson, J. Isoya, and J. Wrachtrup, "Perfect alignment and preferential orientation of nitrogen-vacancy centers during chemical vapor deposition diamond growth on (111) surfaces," Applied Physics Letters **104**, 102407 (2014).

[26] Y. Zhang, S. Yu, L. Zhang, T. Zhang, Y. Yang, and H. Li, "Analysis of the photonic bandgaps for gyrotron devices," Plasma Science, IEEE Transactions on **43**, 1018–1023 (2015).

[27] V. E. Boria and B. Gimeno, "Waveguide filters for satellites," Microwave Magazine, IEEE **8**, 60–70 (2007).

[28] A. Moroz, "A hidden analytic structure of the rabi model," Annals of Physics **340**, 252–266 (2014).

[29] J.-M. Le Floch, C. Bradac, N. Nand, S. Castelletto, M. E. Tobar, and T. Volz, "Addressing a single spin in diamond with a macroscopic dielectric microwave cavity," Applied Physics Letters **105**, 133101 (2014).

[30] S. Probst, A. Tkalčec, H. Rotzinger, D. Rieger, J.-M. Le Floch, M. Goryachev, M. E. Tobar, A. V. Ustinov, and P. A. Bushev, "Three-dimensional cavity quantum electrodynamics with a rare-earth spin ensemble," Phys. Rev. B **90**, 100404 (2014).

[31] W. W. Hansen, "A type of electrical resonator," Journal of Applied Physics **9**, 654–663 (1938)

[32] S. K. Remillard, A. Hardaway, B. Mork, J. Gilliland, and J. Gibbs, "Using a re-entrant microwave resonator to measure and model the dielectric breakdown electric field of gases," Progress In Electromagnetics Research B **15**, 175–195 (2009).

[33] J.-M. Le Floch, Y. Fan, M. Aubourg, D. Cros, N. Carvalho, Q. Shan, J. Bourhill, E. Ivanov, G. Humbert, V. Madrangeas, *et al.*, "Rigorous analysis of highly tunable cylindrical transverse magnetic mode re-entrant cavities," Review of Scientific Instruments **84**, 125114 (2013).

[34] D. L. Creedon, J.-M. Le Floch, M. Goryachev, W. G. Farr, S. Castelletto, and M. E. Tobar, "Strong coupling between $p1$ diamond impurity centers and a three-dimensional lumped photonic microwave cavity," Phys. Rev. B **91**, 140408 (2015).

[35] C. A. Flory and R. C. Taber, "High performance distributed Bragg reflector microwave resonator," Ultrasonics, Ferroelectrics and Frequency Control, IEEE Transactions on **44**, 486–495 (1997).

[36] J.-M. Le Floch, M. E. Tobar, D. Cros, and J. Krupka, "High q-factor distributed Bragg reflector resonators with reflectors of arbitrary thickness," Ultrasonics, Ferroelectrics and Frequency Control, IEEE Transactions on **54**, 2689–2695 (2007).

[37] J.-M. le Floch, M. E. Tobar, D. Mouneyrac, D. Cros, and J. Krupka, "Discovery of Bragg confined hybrid modes with high q factor in a hollow dielectric resonator," Applied Physics Letters **91**, 142907 (2007).

[38] M. E. Tobar, J.-M. Le Floch, D. Cros, J. Krupka, J. D. Anstie, and J. G. Hartnett, "Spherical Bragg reflector resonators," Ultrasonics, Ferroelectrics and Frequency Control, IEEE Transactions on **51**, 1054–1059 (2004).

[39] E. Snitzer, "Cylindrical dielectric waveguide modes," JOSA **51**, 491–498 (1961).

[40] O. Di Monaco, Y. Kersalé, and V. Giordano, "Resonance degeneration and spurious mode suppression in a cryogenic whispering gallery mode sapphire resonator," Microwave and Guided Wave Letters, IEEE **10**, 368–370 (2000).

[41] J.-M. Le Floch, J. D. Anstie, M. E. Tobar, J. G. Hartnett, P.-Y. Bourgeois, and D. Cros, "Whispering modes in anisotropic and isotropic dielectric spherical resonators," Physics Letters A **359**, 1–7 (2006).

[42] J. Sirigiri, K. Kreischer, J. Machuzak, I. Mastovsky, M. Shapiro, and R. Temkin, "Photonic band-gap resonator gyrotron," Physical Review Letters **86**, 5628–5631 (2001).

[43] H. Altug and J. Vukovi, "Two-dimensional coupled photonic crystal resonator arrays," Applied Physics Letters **84**, 161–163 (2004).

[44] G. Humbert, J.-M. Le Floch, D. Mouneyrac, D. Frachou, M. Aubourg, M. Tobar, D. Cros, and J.-M. Blondy, "Hollow-core resonator based on out-of-plane two-dimensional photonic band-gap crystal cladding at microwave frequencies," Applied Physics Letters **96** (2010).

[45] D. Férachou, G. Humbert, J.-M. Le Floch, M. Aubourg, J.-L. Auguste, M. Tobar, D. Cros, and J.-M. Blondy, "Compact hollow-core photonic band gap resonator with optimised metallic cavity at microwave frequencies," Electronics Letters **47**, 805–807 (2011).

[46] Y. Zhang, S. Yu, L. Zhang, T. Zhang, Y. Yang, and H. Li, "Analysis of the photonic bandgaps for gyrotron devices," IEEE Transactions on Plasma Science (2015).

[47] J. Krupka, K. Derzakowski, M. Tobar, J. Hartnett, and R. G. Geyer, "Complex permittivity of some ultralow loss dielectric crystals at cryogenic temperatures," Measurement Science and Technology **10**, 387 (1999).

[48] A. Luiten, M. Tobar, J. Krupka, R. Woode, E. Ivanov, and A. Mann, "Microwave properties of a rutile resonator between 2 and 10 k," Journal of Physics D: Applied Physics **31**, 1383 (1998).

[49] J.-M. Le Floch, M. E. Tobar, D. Cros, and J. Krupka, "Low-loss materials for high q-factor Bragg reflector resonators," Applied Physics Letters **92**, 032901 (2008).

[50] M. D. Janezic and J. Baker-Jarvis, "Full-wave analysis of a split-cylinder resonator for nondestructive permittivity measurements," Microwave Theory and Techniques, IEEE Transactions on **47**, 2014–2020 (1999).

[51] Q. Simon, V. Bouquet, W. Peng, J.-M. Le Floch, F. Houdonougbo, S. Dputier, S. Weber, A. Dauscher, V. Madrangeas, D. Cros, and M. Guilloux-Viry, "Reduction of microwave dielectric losses in kta1 - xnbxo3 thin films by mgo-doping," Thin Solid Films **517**, 5940–5942 (2009).

[52] J.-M. Le Floch, F. Houndonougbo, V. Madrangeas, D. Cros, M. Guilloux-Viry, and W. Peng, "Thin film materials characterization using TE modes cavity," Journal of Electromagnetic Waves and Applications **23**, 549–559 (2009).

[53] R. Chisholm, "Attenuation in wedge and septate waveguides," Antennas and Propagation, IRE Transactions on **7**, 279–283 (1959).

[54] D. Budimir and C. W. Turner, "Novel high-q waveguide e-plane resonators using periodic metallic septa," Microwave and Optical Technology Letters **23**, 311–312 (1999).

[55] N. N. Esfahani, P. Rezaee, K. Schünemann, R. Knöchel, and M. Tayarani, "Miniaturized coaxial cylindrical cavity filters based on sub-wavelength metamaterial loaded resonator," in *Electromagnetics in Advanced Applications (ICEAA), 2011 International Conference on* (IEEE, 2011 pp. 1086–1089.

[56] J.-M. Le Floch, R. Bara, J. G. Hartnett, M. E. Tobar, D. Mouneyrac, D. Passerieux, D. Cros, J. Krupka, P. Goy, and S. Caroopen, "Electromagnetic properties of polycrystalline diamond from 35 k to room temperature and microwave to terahertz frequencies," Journal of Applied Physics **109**, 094103 (2011).

[57] D. L. Creedon, M. E. Tobar, J.-M. Le Floch, Y. Reshitnyk, and T. Duty, "Single-crystal sapphire resonator at millikelvin temperatures: Observation of thermal bistability in high-q factor whispering gallery modes," Physical Review B **82**, 104305 (2010).

[58] J. D. Anstie, J. G. Hartnett, M. E. Tobar, J. Winterflood, D. Cros, and J. Krupka, "Characterization of a spherically symmetric fused-silica-loaded cavity microwave resonator," Measurement Science and Technology **14**, 286 (2003).

[59] J. Molla, A. Ibarra, J. Margineda, J. Zamarro, and A. Hernandez, "Dielectric property measurement system at cryogenic temperature and microwave frequencies," Instrumentation and Measurement, IEEE Transactions on **42**, 817–821 (1993).

[60] M. Goryachev, W. G. Farr, D. L. Creedon, Y. Fan, M. Kostylev, and M. E. Tobar, "High-cooperativity cavity qed with magnons at microwave frequencies," Physical Review Applied **2**, 054002 (2014).

[61] S. R. Parker, J. G. Hartnett, R. G. Povey, and M. E. Tobar, "Cryogenic resonant microwave cavity searches for hidden sector photons," Phys. Rev. D **88**, 112004 (2013).

[62] W. F. Koehl, B. B. Buckley, F. J. Heremans, G. Calusine, and D. D. Awschalom, "Room temperature coherent control of defect spin qubits in silicon carbide," Nature **479**, 84–87 (2011).

[63] H. Kraus, V. Soltamov, D. Riedel, S. Väth, F. Fuchs, A. Sperlich, P. Baranov, V. Dyakonov, and G. Astakhov, "Room-temperature quantum microwave emitters based on spin defects in silicon carbide," Nature Physics **10**, 157–162 (2014).

[64] S. Castelletto, B. Johnson, V. Ivády, N. Stavrias, T. Umeda, A. Gali, and T. Ohshima, "A silicon carbide room-temperature single-photon source," Nature materials **13**, 151–156 (2014).

[65] P. Siyushev, K. Xia, R. Reuter, M. Jamali, N. Zhao, N. Yang, C. Duan, N. Kukharchyk, A. Wieck, R. Kolesov, *et al.*, "Coherent properties of single rare-earth spin qubits," Nature communications **5** (2014).